\documentclass[prb,aps,twocolumn,showpacs]{revtex4}
\usepackage{graphicx}
\usepackage{dcolumn}
\usepackage{bm}
\usepackage{color}
\begin{document}

\title{Stationary phase approximation approach to the quasiparticle interference
on the surface of a strong topological insulator}
\author{Qin Liu $^{1,2}$}
\author{Xiao-Liang Qi $^{2}$}
\author{Shou-Cheng Zhang $^{2}$}
\affiliation{$^1$ State Key Laboratory of
Functional Materials for Informatics, Shanghai Institute of
Microsystem and Information Technology, CAS, Shanghai 200050, China}
\affiliation{$^2$ Department of Physics, Stanford University, Stanford, California 94305, USA}

\date{\today}

\begin{abstract}
Topological insulators have surface states with unique spin-orbit coupling.
With impurities on the surface, the quasiparticle interference pattern is an effective way to
reveal the topological nature of the surface states, which can be probed by the scanning tunneling
microscopy. In this paper, we present a general analytic formulation of the
local density of states using the stationary phase approximation.
The power laws of Friedel oscillations are discussed for a constant energy contour with a generic shape.
In particular, we predict unique signature of magnetic impurities in comparison
with nonmagnetic impurities for a surface state trapped in a ``magnetic wall''.
\end{abstract}

\maketitle

\section{Introduction}
Topological insulators in three dimensions (3D) are band insulators which have a
bulk insulating gap and gapless surface
states with odd number of Dirac cones protected by time-reversal
symmetry (TRS). \cite{Qi2011RMP,Hasan2010,Moore2010} A family of 3D
topological insulators (TI) with a large bulk gap and a single Dirac cone on the
surface includes the compounds Bi$_2$Se$_3$, Bi$_2$Te$_3$ and Sb$_2$Te$_3$,
which have been theoretically predicted and experimentally observed. \cite{Zhang2009,Xia2009,Chen2009,Hsieh2009}
The surface state of these materials can be described by the effective Dirac Hamiltonian
$H_0=\hbar v_F\hat z\cdot(\mbox{\boldmath$\sigma$}\times
{\bf k})$ (with ${\bf k}=(k_x,k_y)$ the momentum) when the Fermi level is close to the Dirac point,
which behaves like a massless relativistic Dirac fermion with the spin locked to its
momentum.\cite{raghu2010} However, compared to the familiar Dirac fermions in particle physics, those
emergent quasiparticles exhibit richer behaviors.
In Bi$_2$Te$_3$, an unconventional hexagonal warping effect
appears due to the crystal symmetry, \cite{Fu2009} which means the constant energy contour (CEC)
of the surface band evolves from a convex circle to a concave hexagon
as the energy moves away from the Dirac point. Although the topological property
of the surface states is not affected, such kind of deformation of the CEC
does affect the behavior of the surface states in the presence of impurities.

Quasiparticle interference (QPI) caused by impurity scattering on the surface
of 3DTIs is an effective way to reveal the topological nature of the surface states.
The interference between incoming and outgoing waves at momenta ${\bf k}_i$ and ${\bf k}_f$
leads to an amplitude modulation of the local density of state (LDOS) at wave vector
${\bf q}={\bf k}_f-{\bf k}_i$, known as the Friedel oscillation. \cite{Friedel1952}
Nowadays, such modulation can be studied by a powerful
surface probe, the scanning tunneling microscopy (STM), which directly measures the LDOS.
The information in momentum space is obtained through Fourier transform scanning tunneling spectroscopy (FT-STS).
Several STM measurements \cite{Yazdani2009,Manoharan2009,Alpichshev2009,Xue2009,hanaguri2010,Wang2011,alpichshev2011,alpichshev2011b,beidenkopf2011}
have been performed on the surface of 3DTIs in the presence of nonmagnetic point and edge impurities,
and the following features are shared in common.
(i) The topological suppression of backward scattering from nonmagnetic point and
edge impurities is confirmed by the observation of strongly damped oscillations in LDOS,
together with the invisibility of the corresponding scattering wave vector ${\bf q}$ in FT-STS.
(ii) Anomalous oscillations are reported in Bi$_2$Te$_3$
for both point and edge impurities when the CEC becomes concave.
These experimental facts have been interpreted theoretically
by several groups. \cite{Zhou2009,Lee2009,Guo2010,Biswas,Wang2010,Wang2011}
For short-range point and edge impurities,
the Friedel oscillation in an ordinary two-dimensional electron gas (2DEG) has the power
law of $R^{-1}$ and $R^{-1/2}$ respectively. \cite{Crommie1993} In comparison,
the Friedel oscillation in a helical liquid with a convex CEC is dominated by the
scattering between time-reversed points (TRP) and is thus suppressed to
$R^{-2}$ and $R^{-3/2}$ for point and edge impurities separately. This result is the crucial reason of
the invisibility of the scattering wave vector ${\bf q}$ in FT-STS,
and is the direct consequence of the suppression of backscattering protected
by TRS in helical liquid. When the CEC becomes concave, scattering between wave vectors,
which are not connected by TRP, can have a significant contribution and
leads to a slower decay of the Friedel oscillation.\cite{Fu2009,Alpichshev2009}

Motivated by these results, in this work we develop a general theory of the QPI for
a CEC of generic shape using the stationary phase approximation approach. \cite{Ruth1966}
This approach has been applied successfully to the Ruderman-Kittel-Kasuya-Yosida
interaction in 3D systems with nonspherical Fermi surfaces. \cite{Ruth1966}
In the stationary phase approximation, the long distance behavior of the
Friedel oscillations is dominated by the so-called ``stationary points'' on the CEC.
Using this approach, a complete result of the power-expansion series of the LDOS and spin LDOS is obtained
for both point- and edge-shaped nonmagnetic and magnetic impurities,
which we model by $\delta$-function potentials. The spin LDOS is the
local spin density at a given energy, which can be measured by a STM experiment with a magnetic tip.
Our results depend only on the TRS and the local geometry around the stationary points
on the CEC, which explain not only the usual
$R^{-1}$ and $R^{-1/2}$ power laws in 2DEG but also the $R^{-2}$ and $R^{-3/2}$
oscillations in the helical liquid. With a generic shape of CEC,
a different power law can be obtained due to the presence of additional stationary points besides the TRP,
which can be used to predict the result of STM and spin-resolved STM experiments
on the surface of other TI materials with more complicated surface states.
An important consequence of our result is that an ordinary STM measurement
cannot distinguish magnetic and nonmagnetic impurities, although the former can
induce backscattering while the latter cannot. To distinguish the effect of magnetic
and nonmagnetic impurities and observe backscattering induced by magnetic impurities,
it is necessary to use a magnetic tip to measure the spin LDOS.

The rest of this paper is organized as follows. In Sect. \ref{standingwave}, we
introduce an intuitive picture of the interference between
helical waves scattered by magnetic impurities.
In Sect. \ref{general formulation}, we
present the general analytic formulation of LDOS for point and
edge impurities respectively by focusing firstly on
those CEC where the stationary points are extremal points. We then generalize our
results to the more generic CEC where the stationary points are saddle points,
with the first nonzero expansion coefficient occurring at a higher power.
Conclusion and discussion are given in Sect. \ref{conclusion}.

\section{Standing wave of the spin interference between two helical waves}
\label{standingwave}
\begin{figure}[tbp]
\begin{center}
\includegraphics[width=2.8in]{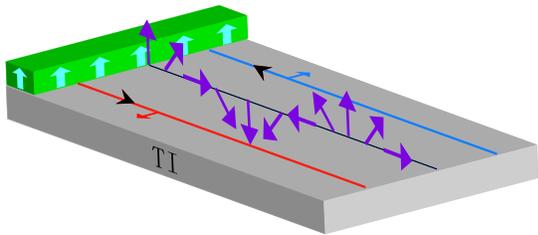}
\end{center}
\caption{(Color online) Illustration of charge and spin interference patterns between two
counter-propagating helical waves.
The gray block is a 3DTI with a magnetic
edge impurity (green stripe) lying along the $x$-axis on the surface.
An incident helical wave along $y$-direction with spin
polarized in the $x$-direction (blue line) is backscattered by the
magnetic edge and the spin is flipped (red line).
The interference of the two orthogonal helical waves leads to
a constant LDOS in the charge channel,
but a spiral LDOS in the spin channel (purple arrows) in the $yz$-plane.}
\label{fig1}
\end{figure}

With the presence of TRS, the backscattering by nonmagnetic impurities
is known to be forbidden on the surface of 3DTIs, due to the $\pi$ Berry's phase
associated with the full rotation of electron spin. \cite{Ando1998,Qi2010}
In experiments, this manifests in the invisibility of the scattering
wave vector $2k_F$ in FT-STS. \cite{Yazdani2009}
It would then be interesting to ask how the surface states
respond differently to magnetic impurities, and what are their
characteristic signatures in STM measurements. With magnetic impurities,
naively one would expect to see a nontrivial interference pattern since
backscattering is allowed due to the breaking of TRS.
However, it turns out that the Friedel oscillation in the charge LDOS,
which is measured in an ordinary STM experiment with a nonmagnetic tip,
is still suppressed in the same way as nonmagnetic impurities.
The broken TRS would only manifest itself in the spin LDOS measured by
a spin-polarized STM tip. \cite{Liu2009}

To understand this result, we first present a simple picture of the interference between two
counter-propagating helical waves on the surface of a 3DTI, and then
give a complete theoretical survey in the next section.
Consider a magnetic edge impurity placed along the $x$-axis
on the surface. For the effective Hamiltonian $H_0=\hbar v_F\hat z\cdot(\mbox{\boldmath$\sigma$}\times
{\bf k})$, the electron state propagating along $y$ direction perpendicular
to the impurity line has spin polarized to $x$-direction, with the wavefunction
$\psi_1=\frac{1}{\sqrt{2}}e^{ik_Fy}(1\;1)^{\text {T}}$. Here the superscript ``T'' indicates the transpose.
This wave is then backscattered by the magnetic edge and counter-propagates in
$-y$-direction. For the same energy, the state with opposite ${\bf k}$ must have opposite spin,
with the wavefunction $\psi_2=\frac{1}{\sqrt{2}}e^{-ik_Fy}(-1\;1)^{\text T}$. This situation is illustrated in Fig.\ref{fig1}.
A simple calculation shows that the interference of the
two counter-propagating helical waves, $\psi(y)=\frac{1}{\sqrt{2}}(\psi_1(y)+\psi_2(y))$, leads to a constant
charge LDOS on the surface $\left\langle \rho\right\rangle_\psi=|\psi^\dagger \psi(y)|=1$
since $\psi_1$ and $\psi_2$ have orthogonal spin. However the interference leads to a spiral spin LDOS in $yz$-plane as
$\langle{\bf s}\rangle_{\psi}=\psi^{\dagger}{\bf s}\psi=[0, -\frac{\hbar}{4}\sin(2k_Fy),-\frac{\hbar}{4}\cos(2k_Fy)
]$, where ${\bf s}=\frac{\hbar}{2}\mbox{\boldmath$\sigma$}$ is the electron spin
operator. Therefore a STM experiment with a nonmagnetic tip will observe no interference pattern
while one with a magnetic tip will observe the oscillation of the spin density of states.
Such a contrast between charge and spin density of states is a unique signature of the helical liquid,
which is a direct demonstration of the locking between spin and momentum.

\begin{figure}[tbp]
\includegraphics[width=2.4in]{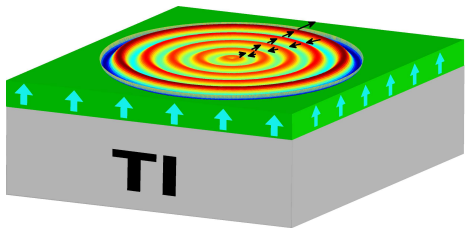}
\caption{(Color online) Standing wave of spin interference
between two helical waves inside a closed ``magnetic wall''
on top of a 3DTI surface. The magnetic wall is surrounded by
a magnetic layer deposited on top of the 3DTI surface,
which opens a gap in the helical surface states and plays the role of a barrier.
The out-of-plane spin LDOS is exhibited by the colored rings, and the
in-plane spin LDOS is indicated by the dark arrows.}
\label{fig2}
\end{figure}

To observe such a spin interference pattern, a more convenient setup is a
closed ``magnetic wall'' as shown in Fig.\ref{fig2}.
Consider a magnetic layer deposited everywhere on the 3DTI surface except a hole in the middle with the disk shape.
The magnetic layer can open a gap on the surface state,
such that the low energy surface states are trapped in the hole region and form standing wave.
Similar to the straight line magnetic impurity discussed above, the standing wave trapped by the magnetic barrier can be obtained by setting the boundary condition of fixed spin at the boundary of the hole. For large $R$ ($R\gg 1/k_F$) the spin density of the standing wave has the behavior of
$\langle s_R\rangle\sim \frac{\sin(2k_FR)}{\sqrt{R}}$,
$\langle s_z\rangle\sim\frac{\cos(2k_FR)}{\sqrt{R}}$, with $R$ and $z$ standing for
longitudinal and perpendicular directions in a spherical coordinate. A unique property of the helical surface states is the spin-charge locking\cite{raghu2010}. For the effective Hamiltonian $H_0=\hbar v_F\hat z\cdot(\mbox{\boldmath$\sigma$}\times
{\bf k})$, the electric current operator in the long wavelength limit is ${\bf j}=\nabla_{\bf k} H_0=\hbar v_F\hat{z}\times \mbox{\boldmath$\sigma$}$. Therefore there is a
loop charge current $j_{\phi}=-2e v_F\langle s_R\rangle$ along the azimuthal direction
associated with the spin density.

\section {General formulation of Stationary phase approximation approach to QPI on the surface of 3DTI}
\label{general formulation}
In this section, we obtain the general long-distance features of charge
and spin LDOS on the surface of a generic 3DTI induced by nonmagnetic and
magnetic impurities using the stationary phase approximation method. \cite{Ruth1966}
We study both point-like and edge-like impurities.
We shall focus first on the behavior of a special kind of CEC where the stationary
points are extremal points,
and then generalize our results to generic CEC with higher order nesting points.

\subsection{Point impurity}
\label{pointimpurity}
We start by considering a point defect on the surface of a 3DTI.
The Hamiltonian with a single impurity is
\begin{eqnarray}
H=\int d^2{\bf r}\psi^\dagger({\bf r})\left[h_0({\bf k}) +V\sigma^\mu\delta({\bf r})\right]\psi({\bf r}),
\end{eqnarray}
where $\sigma^\mu=1$ for $\mu=0$ and $\sigma^a=\sigma^{x,y,z}$ is the Pauli matrices
for $a=1,2,3$. ${\bf k}=-i\nabla$ is the momentum operator.
For such a potential the LDOS can be expressed exactly. Using the $\sigma^\mu$ matrices,
the charge and spin LDOS are combined to the form
\begin{eqnarray}
\rho_\nu(\omega,{\bf R})=-\frac{1}{\pi}{\rm Im}\{{\rm tr}\left[G^r(\omega,{\bf R},{\bf R})\sigma^\nu\right]\},
\end{eqnarray}
with $G^r(\omega,{\bf R},{\bf R^{\prime}})$ being the retarded Green's function in real space.
Let $\rho_{\nu 0}(\omega)$ be the LDOS of the unperturbed system with $V=0$, the deviation of the LDOS
from the background value $\rho_{\nu 0}(\omega)$ is then given by
\begin{eqnarray}
\delta\rho_{\mu\nu}(\omega,{\bf
R})&\equiv &\rho_\nu(\omega,{\bf R})-\rho_{\nu0}(\omega)\nonumber\\
&=&-\frac{1}{\pi}\text{Im}\int\frac{d^2kd^2k^{\prime}}{(2\pi)^4}e^{i({\bf
k-k^{\prime}})\cdot{\bf R}}\times\nonumber\\
&&\text{tr}\left[{G}^r_0(\omega,{\bf
k})T^{\mu}(\omega,{\bf k},{\bf k}^{\prime}){G}^r_0(\omega,{\bf
k^{\prime}})\sigma^{\nu}\right].\label{ldosptdef}
\end{eqnarray}
Here $G^r_0(\omega,{\bf k})$ is the free retarded Green's function
governing the CEC under consideration. For the topological surface states $G^r_0(\omega,{\bf k})=(\omega+i\delta-h_0({\bf k}))^{-1}$.
The T-matrix $T^{\mu}(\omega,{\bf k,k'})$ is defined by
\begin{eqnarray}
T^{\mu}(\omega)=V\sigma^{\mu}\left[1-V\sigma^{\mu}G^r_0(\omega)\right]^{-1},
\end{eqnarray}
which is momentum independent when the impurity has a $\delta$-function potential,
and we have denoted the real space Green's function
$G^r_0(\omega)=\int \frac{d^2k}{(2\pi)^2}{G}^r_0(\omega,{\bf k})$.
As is required by the TRS, $G_0^r(\omega)$ is always proportional to the identity matrix.

We first note that the spin LDOS induced by a nonmagnetic impurity
vanishes uniformly, {\it i.e.}, $\delta\rho_{0a}\equiv 0$ for $a=1,2,3$.
This is a direct consequence of TRS, because under time-reversal transformation $\Theta=i\sigma^y$,
we have $\Theta^{-1}\sigma^a\Theta=-\sigma^{a^{\text T}}$ and $\Theta^{-1}G^r_{0,k}\Theta=G_{0,-k}^{r^{\text T}}$,
then the trace in Eq.(\ref{ldosptdef}) satisfies
$\text {tr}[G^r_{0,k}T^0G^r_{0,k^{\prime}}\sigma^a]=-\text {tr}[G^r_{0,-k^{\prime}}T^0G^r_{0,-k}\sigma^a]$,
where we have abbreviated $G^r_0(\omega,{\bf k})\equiv G^r_{0,k}$.
By interchanging ${\bf k}$ and $-{\bf k}^{\prime}$ in the integral in Eq.(\ref{ldosptdef}),
one is led to the result $\delta\rho_{0a}(\omega,{\bf R})=0$.
To obtain other components of the T-matrix, we
expand the T-matrix into a spin-dependent and spin-independent parts as
\begin{eqnarray}
T^a&=&T^a_a\sigma^a+T^a_0\nonumber\\
T^a_a&=&\frac V{1-V^2G^{r2}_0(\omega)},~T^a_0=\frac{V^2G^r_0(\omega)}{1-V^2G^{r2}_0(\omega)}\nonumber\\
T^0&=&\frac V{1-VG^{r}_0(\omega)}
\end{eqnarray}
where the fact that $G^r_0(\omega)$ is proportional to identity has been used,
and no summation over repeated indices is implied throughout the paper.
Similar to the argument in $\delta\rho_{0a}$ case, we see that the contribution of $T^a_a$ to the
charge LDOS of a magnetic impurity $\delta\rho_{a0}$
vanishes. Hence we have $\delta\rho_{a0}/\delta \rho_{00}=T_0^a/T^0$. 
Therefore, in the following, we shall
focus only on $\delta\rho_{00}$ and $\delta\rho_{ab}$.

To proceed, the measured LDOS in Eq.(\ref{ldosptdef}) is then rewritten
in the diagonal basis of the topological surface bands.
Define the unitary matrices $U_{\bf k}$ such that $U_{\bf k}^\dagger h_0({\bf k})U_{{\bf k}}$ diagonalizes $h_0({\bf k})$,
Eq.(\ref{ldosptdef}) becomes
\begin{eqnarray}
\delta\rho_{\mu\nu}(\omega,{\bf
R})=&&-\frac{1}{\pi}\text{Im}\int\frac{d^2kd^2k^{\prime}}{(2\pi)^4}e^{i({\bf
k-k^{\prime}})\cdot{\bf
R}}\text{tr}\left[(U^{\dagger}_kG^r_{0,k}U_k)\right.\nonumber\\
&&\left.(U^{\dagger}_kT^{\mu}U_{k^{\prime}})(U^{\dagger}_{k^{\prime}}G^r_{0,k^{\prime}}U_{k^{\prime}})
(U^{\dagger}_{k^{\prime}}\sigma^{\nu}U_k)\right]\label{eq2}\\
=&&-\frac{1}{\pi}\text{Im}\int\frac{d^2kd^2k^{\prime}}{(2\pi)^4}e^{i({\bf
k-k^{\prime}})\cdot{\bf
R}}\times\nonumber\\
&&\sum_{nm}\frac{\gamma^{\mu\nu}_{nm}({\bf k},{\bf k}^{\prime})
\Sigma^{\nu^{\ast}}_{nm}({\bf k},{\bf k}^{\prime})}
{(\omega+i\delta-\varepsilon_n)(\omega+i\delta-\varepsilon_m^{\prime})},
\label{eq3}
\end{eqnarray}
where $\varepsilon_{n(m)}({\bf k})$ are the energy eigenvalues of the
bands $|n(m){\bf k}\rangle$, and we have defined $\Sigma^{\mu}_{nm}({\bf
k,k^{\prime}})=\langle n{\bf k}|\sigma^{\mu}|m{\bf
k^{\prime}}\rangle$, as well as
\begin{eqnarray}
\gamma^{\mu\nu}_{nm}({\bf k},{\bf k}^{\prime})=
\left\{\begin{array}{cc}
T^0\Sigma^0_{nm},\;\;\;\;\;\;\;\;\;\;\;\;\;\;\mu=\nu=0,\\
T^a_a\Sigma^a_{nm}+T^a_0\Sigma^0_{nm},\; \mu=a,\nu=b.
\end{array}\right.
\end{eqnarray}
Following the standard process of density of states calculations, \cite{Ruth1966}
the integrations over ${\bf k}$ and ${\bf k}^{\prime}$ are then converted into
coordinates $d{\bf k}=(dk_{\bot},dk_{\phi})$ as
\begin{eqnarray}
\delta\rho_{\mu\nu}(\omega,{\bf
R})=&-&\frac{1}{\pi}{\text Im}\oint\frac{dk_{\phi}dk_{\phi}^{\prime}e^{i({\bf
k-k^{\prime}})\cdot{\bf
R}}}{(2\pi)^4}\int\frac{d\varepsilon_nd\varepsilon_m^{\prime}}
{|\nabla_{\bot}\varepsilon_n\nabla^{\prime}_{\bot}\varepsilon_m^{\prime}|}\nonumber\\
&\times&
\sum_{nm}\frac{\gamma^{\mu\nu}_{nm}({\bf k},{\bf k}^{\prime})
\Sigma^{\nu^{\ast}}_{nm}({\bf k},{\bf k}^{\prime})}
{(\omega+i\delta-\varepsilon_n)(\omega+i\delta-\varepsilon_m^{\prime})},
\label{eq5}
\end{eqnarray}
where $k_{\bot}$ and $k_{\phi}$ are components of ${\bf k}$ normal
and tangential to the CEC, respectively.

\begin{figure}[tbp]
\begin{center}
\includegraphics[width=3.8in]{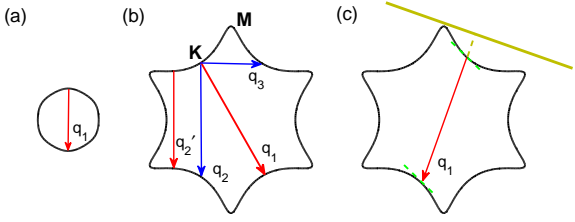}
\end{center}
\caption{(Color online) Schematic picture of CEC and stationary points
for point and edge impurities. (a) Convex CEC where
there is only one pair of stationary points connected
by the red arrow along any given direction for both point and line
impurities. (b) Concave CEC for point impurity
where there are multiple pairs of stationary points. Nonstationary
points are shown for example as blue arrows. (c)
Concave CEC for edge impurities (brown line) where the slopes (green dashed lines)
at the pair of stationary points are the same.}
\label{fig3}
\end{figure}
To evaluate the loop integrals along the CEC, it is essential to introduce
the stationary phase approximation. For example, consider the LDOS at a point ${\bf R}=R\hat y$
(here and hereafter we shall always take the $y$-direction for example),
the phase factor $e^{i{\bf (k-k')\cdot R}}=e^{i(k_y-k_y')R}$. Locally,
one can write $k_y=k_y(\varepsilon, k_x)$ as a function of energy $\varepsilon$ and $k_x$.
For large distance $R$ from the impurity, the phase factors
$e^{ik_y(\varepsilon,k_x)R}$ and $e^{-ik_y^{\prime}(\varepsilon^{\prime},k_x^{\prime})R}$
vary rapidly with respect to $k_x$ and $k_x^{\prime}$ for almost every point on the CEC, so that most of the
integrations cancel out exactly except for the stationary points, ${\bf k}_i$, \cite{Ruth1966}
which satisfy the condition
\begin{eqnarray}
\frac{\partial k_y(\varepsilon,k_x)}{\partial k_x}=
\frac{\partial k_y^{\prime}(\varepsilon^{\prime},k_x^{\prime})}{\partial k_x^{\prime}}=0.
\label{pointcnd}
\end{eqnarray}
The stationary points defined above include i) extremal points such as the pairs connected by ${\bf q}_1$
in Figs.\ref{fig3}(a) and (b), where the second derivative $\partial^2k_y/\partial k_x^2$ is nonvanished;
ii) the turning points such as the pair connected by ${\bf q}_2^{\prime}$
in Fig.\ref{fig3}(b), where the second derivative also vanishes.
In the following, we first focus only on the extremal points,
and leave the more general discussions to Sect.\ref{arbitraryCEC}.

Having identified the pairs of stationary points on the CEC
in direction ${\bf R}$, the loop integrals in Eq.(\ref{eq5})
at large distances are then approximated
by the summation of integrals in the neighborhood of all the stationary-point
pairs, which is the essence of the method of the stationary phase approximation.
To start with, we first change the integral
variables as $d^2k=d\varepsilon dk_x/\hbar|v_{yi}|$,
where $v_{yi}=\partial
\varepsilon({\bf k})/\hbar\partial k_{yi}$, and then expand the
CEC at the extremal points as $k_y=k_{yi}-(k_x-k_{xi})^2/2\rho_{xi}$,
where $\rho_{xi}=-[\partial^2k_{yi}(\varepsilon,k_x)/\partial^2 k_{xi}]^{-1}$
is the principle radii of curvature of the CEC at
the extremal points, which is positive for maxima while negative for minima.
Under this approximation, Eq.(\ref{eq5}) becomes
\begin{eqnarray}
\delta\rho_{\mu\nu}(\omega,{\bf R})\simeq&&
-\frac{1}{\pi}\text{Im}\sum_{mn}\sum_{ij}\int\frac{d\varepsilon_n}{(2\pi)^2}
\frac{1}{\omega+i\delta-\varepsilon_n}\frac{e^{ik_{yi}R}}{\hbar|v_{yi}|}\nonumber\\
&&\times\int\frac{d\varepsilon_m^{\prime}}{(2\pi)^2}
\frac{1}{\omega+i\delta-\varepsilon_m^{\prime}}\frac{e^{-ik_{yj}^{\prime}R}}{\hbar|v_{yj}^{\prime}|}\nonumber\\
&&\times\int_{-\infty}^{\infty}dx\;e^{-i\frac{x^2}{2\rho_{xi}}R}
\int_{-\infty}^{\infty}dx^{\prime}\;e^{i\frac{x^{{\prime}^2}}{2\rho_{xj}^{\prime}}R}\nonumber\\
&&\times\gamma^{\mu\nu}_{nm}({\bf
k},{\bf k}^{\prime})\Sigma^{\nu^{\ast}}_{nm}({\bf k},{\bf k}^{\prime}),
\label{eq7}
\end{eqnarray}
where we have denoted $x=k_x-k_{xi}$, $x^{\prime}=k_x^{\prime}-k_{xj}^{\prime}$, and
all the quantities at the extremal points ($ij$) still depend
on the energies $\varepsilon$ and $\varepsilon^{\prime}$.
Now the matrix element $\Sigma^{\mu}_{nm}({\bf k},{\bf k}^{\prime})$ at the extremal points
is in general some nonzero constant $C^{\mu}_{ni,mj}(\varepsilon,\varepsilon^{\prime})$,
except that it vanishes when $\mu=0$ and the pair of stationary points are time-reversal partners
$|n{\bf k}_i\rangle=\Theta \hat{K}|m {\bf k}^{\prime}_j\rangle$. Here $\hat{K}$ is the complex conjugation operator.
Examples are shown as the pairs of stationary points
connected by ${\bf q}_1$'s in Figs.\ref{fig3}(a) and (b)
for convex and concave CEC respectively. 
To obtain the generic behavior of the interference pattern, the matrix element is expanded in the distance $x,x'$ to the stationary points as
$\Sigma^{\mu}_{nm}(x,x')=C^{\mu}_{ni,mj}+ax+a'x'+o(x)+o(x')$, where $C^{\mu}_{ni,mj}=0$
for $\mu=0$ at TRP, and a nonvanishing but energy dependent constant otherwise.
Inserting the series into Eq.(\ref{eq7}), one can integrate first over $x$
and $x^{\prime}$ by using the relations
$\int_{-\infty}^{\infty}dx e^{iCx^2}=\sqrt{\pi/|C|}e^{i\frac{\pi}{4}\text{sgn}(C)}$
and $\int_{-\infty}^{\infty}dx
x^2e^{iCx^2}=\sqrt{\pi}/(2|C|^{3/2})e^{-i\frac{\pi}{4}\text{sgn}(C)}$,
and then integrate over the energies using the residue theorem
by summation over the integrand at the poles $\varepsilon=\varepsilon^{\prime}=\omega+i\delta$.
Finally by taking the limit $\omega=\varepsilon_F$, $\delta\rightarrow0^{+}$ we get
\begin{eqnarray}
\delta\rho_{\mu\nu}(\omega,{\bf R})&\simeq&
\frac{1}{2\pi^2\hbar^2R}\text{Im}\sum_{mn}\sum_{ij}e^{i(k_{yi}-k_{yj}^{\prime})R}
\frac{|\rho_{xi}\rho_{xj}^{\prime}|^{\frac{1}{2}}}{|v_{yi}v_{yj}^{\prime}|}\nonumber\\
&\times&[e^{i(\phi_i-\phi_j^{\prime})}
\sum_sT^{\mu}_sC^{s}_{ni,mj}C^{\nu^{\ast}}_{ni,mj}+\frac{1}{R}\nonumber\\
&\times&
({a'}^2e^{i(\phi_i+\phi_j^{\prime})}\left|\rho_{xj}^{\prime}\right|
+a^2e^{-i(\phi_i+\phi_j^{\prime})}\left|\rho_{xi}\right|)
]_{\varepsilon_F}\nonumber\\
\label{pointfinal}
\end{eqnarray}
where $\phi_i=-\frac{\pi}{4}\text{sgn}(\rho_{xi})$. This is the long wavelength behavior of LDOS induced by a point impurity.
In the above result, we have $s=0$ and $T^0_0\equiv T^0=V/(1-VG^r_0(\omega))$
for the charge LDOS of a nonmagnetic impurity $\delta\rho_{00}$.
While for the spin LDOS of a magnetic impurity $\delta\rho_{ab}$,
the summation is over $s=a,0$, where $T^a_a$ and $T^a_0$ are respectively the
spin-dependent and spin-independent coefficients in the T-matrix
expansion introduced above.
\begin{figure}[tbp]
\begin{center}
\includegraphics[width=2.8in]{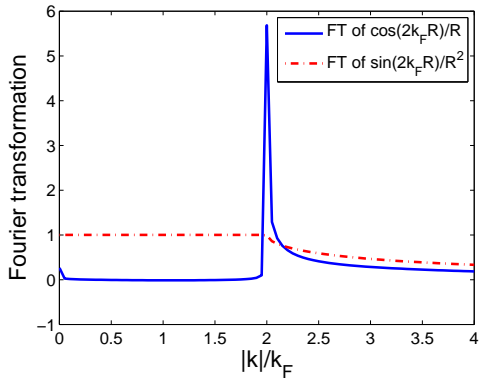}
\end{center}
\caption{(Color online) Fourier transformation of the LDOS with $R^{-1}$ and $R^{-2}$ power laws.}
\label{fig4}
\end{figure}

There are several comments regarding this result.
Firstly, for a pair of non-TRS stationary points like ${\bf q}_2$
in Fig.\ref{fig3}(b), the leading power is given by the
first term in Eq.(\ref{pointfinal}), which is of $R^{-1}$.
While for a pair of TRS stationary points as ${\bf q}_1$
in Figs.\ref{fig3} (a) and (b), the first nonvanishing contribution
to the power law is dominated
by the second term in Eq.(\ref{pointfinal}) as $R^{-2}$ for nonmagnetic impurity,
and for magnetic impurity with ordinary tip.
Such suppression of LDOS is a direct consequence of the absence of backscattering
of helical waves due to TRS. Correspondingly in the fourier transform of LDOS, there is a sharp peak at
$k=2k_F$ for $R^{-1}$ power law, which is absent for $R^{-2}$ power law, as
shown in Fig.\ref{fig4}. For magnetic impurities with
spin-polarized tip, the first term in Eq.(\ref{pointfinal}) dominates
no matter whether the pair of stationary points is TRS or not (due to the contribution of the $s=a$ term),
which gives the visibility of the TRS scattering wave vector ${\bf q}_1$.
This distinct response of surface states to magnetic
impurities from that of nonmagnetic impurities
provides a crucial criteria for the breaking of TRS on the surface of TIs. \cite{Xue2009}
Secondly, in the discussion above we have assumed the matrix element $\Sigma^{\mu}_{nm}$ to be nonzero if it is not forbidden by time-reversal symmetry. There may be some other reasons for the matrix element to vanish. For example, the states at two TRS stationary points have opposite spin. If the impurity spin happens to be parallel (or anti-parallel) to their spin, the matrix element $\Sigma_{nm}^\mu$ can vanish. For non-TRS stationary points, this may occur accidentally, but generically the spin of the two states $n$ and $m$ are not parallel, so that the matrix element is nonvanished for any impurity spin. Since such zeros of matrix elements are at most only realized for some particular directions of the impurity spin, in the following we will focus on the generic cases with nonzero matrix element as long as it is not forbidden by time-reversal symmetry.
Thirdly, in the integral over energy,
we have assumed $v_{yi},v_{yj}^{\prime}\neq 0$ so that the only poles
in the complex energy plane
are $\varepsilon=\varepsilon^{\prime}=\omega+i\delta$.
However, in general, it is possible that there are other poles from
$v_{yi}=0$ or $v_{yj}^{\prime}=0$, which means the stationary points in
CEC are also saddle points in the energy-momentum dispersion.
In that case, we shall further expand $v_{yi}$ (or $v_{yj}^{\prime}$) around $\omega$ as
$v_{yi}(\varepsilon)=v_{yi}(\omega)+(\partial
v_{yi}/\partial
\varepsilon)(\varepsilon-\omega)+\cdots$,
and keep the first nonzero term.
This won't modify the power laws in spatial dependence. \footnote{If the CEC we considered is the Fermi surface, points with zero Fermi velocity may lead to strong effect of electron interaction which may make our discussion invalid. For CEC away from Fermi energy, there is no such effect.}
Finally, note that when summation over the stationary-point
pairs, $(ij)$, we always choose the pair such that one point has
positive velocity $v_{yi}$ and
the other has negative velocity $v_{yj}^{\prime}$. As a summary of the discussion above, the power laws of LDOS
for point impurity are concluded in Table. \ref{pointable}.
\begin{table}[!h]
\tabcolsep 0pt \caption{Power laws of Friedel oscillations for point impurity}
\vspace*{-6pt}
\begin{center}
\def\temptablewidth{0.42\textwidth}
{\rule{\temptablewidth}{1pt}}
\begin{tabular*}{\temptablewidth}{@{\extracolsep{\fill}}ccccccc}
 &  &{ charge LDOS} & spin LDOS \\   \hline
nonmagnetic    &  TRP  & $R^{-2}$ & -  \\
 & non-TRP & $R^{-1}$ &  - \\
magnetic  & TRP &  $R^{-2}$  & $R^{-1}$  \\
&non-TRP & $R^{-1}$ &$R^{-1}$
       \end{tabular*}
       {\rule{\temptablewidth}{1pt}}
       \end{center}
       \label{pointable}
       \end{table}

To provide further intuition on the result (\ref{pointfinal}), we consider some simple examples.
The first example is a 2DES without spin-orbit coupling described by the familiar
 Hamiltonian $H_Q=\hbar^2k^2/2m$, which has two degenerate and isotropic Fermi surfaces, as shown in Fig.\ref{fig5}(a).
According to our theory, the main contribution to the LDOS in this example comes
from the intraband scattering of the same spin orientation between two extremal points,
which we denote as `1' and `2'.
At these points we have
$k_{y2}=\rho_{x2}=k_{\varepsilon}$, $k_{y1}^{\prime}=\rho_{x1}^{\prime}=-k_{\varepsilon^{\prime}}$,
$k_{\varepsilon}=(2m\varepsilon/\hbar^2)^{1/2}$, $v_{y2}=\hbar k_{y2}/m$,
$v_{y1}^{\prime}=\hbar k_{y1}^{\prime}/m$, and $C^0_{11}=C^0_{22}=1$.
Inserting these quantities into Eq.(\ref{pointfinal})
and keeping only to the first order expansion of T-matrix,
we get $\delta\rho^{(1)}_{00}(\omega,R\hat y)\simeq-(Vm^2/\pi^2\hbar^4q)\cos(2qR)/R$,
which has $R^{-1}$ power law. Note that the interband contribution
to the LDOS in this example is from a pair of TRS extremal points,
which has a $R^{-2}$ power law. In contrast, in the example of a 2D Dirac CEC,
$H_D=\gamma\hat{z}\cdot(\mbox{\boldmath$\sigma$}\times{\bf k})$,
there is only one non-degenerate band at a given energy due to the spin splitting,
as shown in Fig.\ref{fig5}(b).
Thus only intraband scattering between a pair of extremal
TRP contributes to the LDOS, and $C^0_{ni,mj}=0$. 
Inserting the quantities $k_{y2}=\rho_{x2}=\varepsilon/\gamma$,
$k_{y1}^{\prime}=\rho_{x1}^{\prime}=-\varepsilon/\gamma$,
and $v_{y1(2)}=\gamma\text{sgn}[k_{y1(2)}]/\hbar$ into Eq.(\ref{pointfinal}),
we get $\delta\rho^{(1)}_{00}(\omega,R\hat y)\simeq(V/4\pi^2\gamma^2)\sin(2qR)/R^2$,
which is consistent with our expectation.

In a recent STM measurement of the TI Bi$_2$Te$_3$ doped with Ag,\cite{Xue2009} clear standing waves and scattering
wave vectors are imaged through
FT-STS when the Fermi surface is of hexagram shape. It is
observed that the high intensity regions are always along the
$\bar\Gamma$-$\bar M$ direction, but the intensity in $\bar\Gamma$-$\bar K$
direction vanishes. This observation can be well-understood
using our stationary phase approximation theory. Among the wave vectors
${\bf q}_1$, ${\bf q}_2$, ${\bf q}_2^{\prime}$ and ${\bf q}_3$ 
 shown in Fig.\ref{fig3}(b), ${\bf q}_1$ and ${\bf q}_2^{\prime}$ correspond to scattering between stationary points,
while ${\bf q}_3$ and ${\bf q}_2$ do not. This explains why no standing waves corresponding
to ${\bf q}_3$ are observed in FT-STS. Within the other two, stationary points
connected by ${\bf q}_1$ are also TRP which shall contribute the power law of $R^{-2}$ according
to our result. Therefore its intensity in FT-STS is too weak to
be observed. For wave vectors ${\bf q}_2$ and  ${\bf q}_2^{\prime}$ along $\bar\Gamma$-$\bar M$ direction,
${\bf q}_2^{\prime}$ is stationary but non-TRS. Our result
shows that this wave vector contributes an $R^{-1}$ power
law, which is responsible for the high intensity reported in Ref.\cite{Xue2009}.

\subsection{Edge impurity}
\label{edgeimpurity}

Beside point impurities, one-dimensional line defect in the form of step edge
has also been observed on the surface of 3DTI. \cite{Manoharan2009,Alpichshev2009}
Magnetic edge defects can possibly be realized by depositing a magnetic layer on top of a 3DTI.
In this section, we discuss the interference
patterns of electronic waves induced by magnetic and nonmagnetic edge defects.

We consider an edge defect along the
$x$-direction on top of a 3DTI surface with the Hamiltonian
$V({\bf r})=V\delta(y)\sigma^{\mu}$. A magnetic edge defect has been illustrated in Fig.\ref{fig1}.
The main difference between an edge defect and a point defect is the momentum conservation along
the edge impurity orientation, which means one
of the loop integrations in Eq.(\ref{eq5}) should be removed. Following similar calculations
as performed in the case of a point impurity, the LDOS for the edge impurity is given by
\begin{eqnarray}
\delta\rho_{\mu\nu}(\omega,{\bf
R})=&-&\frac{1}{\pi}\text{Im}\int\frac{d^2kd^2k^{\prime}}{(2\pi)^4}\delta_{k_x,k_x^{\prime}}e^{i({\bf
k-k^{\prime}})\cdot{\bf R}}\nonumber\\
&\times&\text{tr}\left[{G}^r_0(\omega,{\bf
k})T^{\mu}(\omega,k_x){G}^r_0(\omega,{\bf
k^{\prime}})\sigma^{\nu}\right],\label{ldoslinedef}
\end{eqnarray}
where $T^{\mu}(\omega,k_x)=V\sigma^{\mu}/(1-V\sigma^{\mu} G^r_0(\omega,k_x))$ with
$G^r_0(\omega,k_x)=\int \frac{dk_y}{2\pi} G^r_0(\omega,{\bf k})$.
Similarly as the case of a magnetic point impurity, the T-matrix for
a magnetic edge impurity can again be separated into
a spin-dependent and a spin-independent terms. However, in the following discussion, we shall keep
only to the first order expansion of the T-matrix, $V\sigma^{\mu}$,
which is spin-dependent. This simplification is
appropriate for weak impurity potential, and it
won't affect the qualitative conclusion of the Friedel oscillation power laws,
as we have learned from the case of point impurities.

In the presence of edge impurity, we are usually interested in the
LDOS in the direction perpendicular to the edge orientation.
Similarly as the case of point impurity, the LDOS in eq.(\ref{ldoslinedef})
is first transformed into the diagonal basis of the topological surface bands,
and then converted into integrations over normal and tangential components as in Eq.(\ref{eq5}).
By using the stationary phase approximation,
now the main contribution to the loop integrals comes from such
stationary points where their momentum transfer
${\bf q}$ is normal to the edge orientation,
and the ``slopes'' of CEC at the two stationary points are the same:
\begin{eqnarray}
\frac{\partial}{\partial k_x}\left[k_y(\varepsilon,k_x)
-k_y^{\prime}(\varepsilon^{\prime},k_x)\right]=0. \label{linecnd}
\end{eqnarray}
Compared with the stationary-point condition for point impurity,
the condition for edge impurity allows more possibilities.
One such example is shown schematically as ${\bf q}_1$ in Fig.\ref{fig3}(c)
where the pair of stationary points has the same nonvanished slope.
Such a pair of scattering end
points is not considered as stationary points in the case of point impurities,
but are stationary for edge impurities.
Following the same logic as the discussion
of point impurity in the last section, the CEC is then expanded around the stationary points as
$k_y=k_{yi}+\alpha_{i}(k_x-k_{xi})-(k_x-k_{xi})^2/2\rho_{xi}$,
and the LDOS is approximated by
\begin{widetext}
\begin{eqnarray}
\delta\rho^{(1)}_{\mu\nu}(\omega,{\bf R})\simeq
&&-\frac{V}{\pi}\text{Im}\sum_{mn}\sum_{ij}\int\frac{d\varepsilon_n}{(2\pi)^2}
\frac{1}{\omega+i\delta-\varepsilon_n}\frac{e^{ik_{yi}R}}{\hbar|v_{yi}|}
\int\frac{d\varepsilon_m^{\prime}}{(2\pi)^2}
\frac{1}{\omega+i\delta-\varepsilon_m^{\prime}}\frac{e^{-ik_{yj}^{\prime}R}}{\hbar|v_{yj}^{\prime}|}\nonumber\\
&&\times\int_{-\infty}^{\infty}dxe^{-i\frac{x^2}{2\rho_{xi}}R}
\int_{-\infty}^{\infty}dx^{\prime}e^{i\frac{x^{{\prime}^2}}{2\rho_{xj}^{\prime}}R}
e^{i\alpha_{i}(x-x^{\prime})}\delta_{x,x^{\prime}}
\left[C^{\mu}_{ni,mj}C^{\nu^{\ast}}_{ni,mj}+(ax+a'x')^2\right].
\label{eq11}
\end{eqnarray}
\end{widetext}
Although Eq.(\ref{eq11}) looks similar to
Eq.(\ref{eq7}) in point impurity case, the definition of stationary points
for edge impurity in Eq.(\ref{linecnd}) is quite different from
that of point impurity. Therefore, a lot more terms should be included
in the summation of stationary-point pairs $(ij)$ here compared with
the point impurity case. By
integrating out $x(x^\prime)$ and energy variables, we get
\begin{eqnarray}
\delta\rho^{(1)}_{\mu\nu}(\omega,{\bf R})&\simeq&
\frac{V}{(2\pi)^2\hbar^2}\sqrt{\frac{2}{\pi
R}}\text{Im}\sum_{mn}\sum_{ij}\frac{|P_{ij}|^{1/2}}{|v_{yi}v_{yj}^{\prime}|}\nonumber\\
&\times&e^{i(k_{yi}-k_{yj}^{\prime})R}
\left[C^{\mu}_{ni,mj}C^{\nu^{\ast}}_{ni,mj}e^{i\Phi_{ij}}\right.\nonumber\\
&+&\left.e^{-i\Phi_{ij}}(a+a')^2P_{ij}/R\right]_{\varepsilon_F},
\label{edgefinal}
\end{eqnarray}
where $P_{ij}=\rho_{xi}\rho_{xj}^{\prime}/(\rho_{xj}^{\prime}-\rho_{xi})$
and $\Phi_{ij}=-\frac{\pi}{4}\text{sgn}(P_{ij})$. In the equation above we have assumed $v_{yi},v_{yj}^{\prime}\neq 0$ and
$\rho_{xi}\neq \rho_{xj}^{\prime}$. In other words, this result is not
applicable to the case where the CEC
near the pair of stationary points is nested to the second order expansion.
If such nesting happens, the quadratic terms in the expansion of CEC near
the stationary points cancel out exactly, and higher orders expansion should be employed.
The power laws of Friedel oscillations for edge impurity
are summarized in Table. \ref{linetable},
which shall be used to explain the STM measurements
about edge impurities. \cite{Manoharan2009,Alpichshev2009}
\begin{table}[!h]
\tabcolsep 0pt \caption{Power laws of Friedel oscillations for edge impurity} \vspace*{-6pt}
\begin{center}
\def\temptablewidth{0.42\textwidth}
{\rule{\temptablewidth}{1pt}}
\begin{tabular*}{\temptablewidth}{@{\extracolsep{\fill}}ccccccc}
 &  & ordinary & spin polarized \\   \hline
nonmagnetic  &  TRP  & $R^{-3/2}$ & -  \\
 & non-TRP & $R^{-1/2}$ &  - \\
magnetic & TRP &  $R^{-3/2}$  & $R^{-1/2}$  \\
&non-TRP & $R^{-1/2}$ &$R^{-1/2}$
       \end{tabular*}
       {\rule{\temptablewidth}{1pt}}
       \end{center}
       \label{linetable}
       \end{table}

To have a feeling of how Eq.(\ref{edgefinal}) works explicitly, again we
apply it to the examples of 2DEG Hamiltonian $H_Q$ and 2D Dirac Hamiltonian $H_D$ discussed previously.
A few lines of calculations yield that
for 2D quadratic dispersion,
$\delta\rho_{00}^{(1)}(\omega,R\hat{y})=(Vm^2/2\pi^2\hbar^4q^{3/2})
\sin(2qR-\frac{\pi}{4})/\sqrt{\pi R}$, which is consistent with the experimental observation
in 2DEG.\cite{Crommie1993} For 2D Dirac fermion,
$\delta\rho_{00}^{(1)}(\omega,R\hat{y})=(V/8\pi^2\gamma^2\sqrt{\pi q})
\sin(2qR+\frac{\pi}{4})/R^{3/2}$, which is a consequence of the absence of backscattering
in helical liquid. Information in reciprocal space can be extracted via FT-STS
similarly to the point-impurity case exhibited in Fig.\ref{fig4}, where a notable sharp peak
is present at $k=2k_F$ for a 2DEG, but is absent for the helical liquid.
\begin{figure}[tbp]
\begin{center}
\includegraphics[width=3.2in]{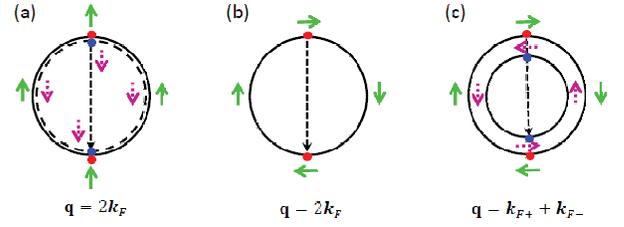}
\end{center}
\caption{(Color online) Schematic CEC of (a) quadratic, (b) Dirac,
and (c) Rashba dispersions. The spin orientations for each degenerate band
are indicated respectively by the green (solid) and purple (dotted)
arrows. The stationary points are represented by red and blue dots,
which are connected by the scattering vector ${\bf q}$ shown as dashed arrows.
The intraband scattering occurs between the stationary points with the same color,
while the interband scattering occurs between those with different colors.}
\label{fig5}
\end{figure}

In an experiment by Gomes {\it et al.},
a nonmagnetic step is imaged by STM topography in Sb $(1 1 1)$ surface.
\cite{Manoharan2009} The Fermi surface consists
of one electron pocket at $\bar\Gamma$ surrounded by six hole pockets
in $\bar\Gamma$-$\bar M$ direction, where the surface dispersion
has a Rashba spin splitting. The measured LDOS in $\bar\Gamma$-$\bar M$ direction is
fitted by a single $q$-parameter using
the zeroth-order of Bessel function of the first kind,
see Fig.2(c) in Ref.\cite{Manoharan2009}, which agrees exactly
with our result in Table. \ref{linetable}. Along $\bar\Gamma$-$\bar M$ direction, the surface
band can be modeled by a Rashba Hamiltonian where the LDOS is dominated
by interband scattering between a pair of non-TRS stationary points,
as shown in Fig.\ref{fig5}(c).
According to our analysis, the Friedel oscillation has $R^{-1/2}$
power law, which is the asymptotic expansion of $J_0(qR)$ at large distances.
Another STM experiment studying the edge impurity
by Alpichshev {\it et. al.} \cite{Alpichshev2009} is in Bi$_2$Te$_3$
where hexagonal warping effect exists, and a nonmagnetic step defect is observed on
crystal surface. A strongly damped oscillation
is reported when the bias voltage is at the energy with
a convex Fermi surface as shown in Fig.\ref{fig3}(a). Though no
fitting of the experimental data is estimated in this region, our
results predict a $R^{-3/2}$ power law. Pronounced oscillations
at higher bias voltages where the hexagon warping effect emerges
are observed with $R^{-1}$ fitting. Despite of the quantitative difference
with our result of $R^{-1/2}$, this $R^{-1}$ oscillation has been
explained in several other works \cite{Zhou2009,Lee2009} beyond
our simple model.

The results summarized in Tables. \ref{pointable} and
\ref{linetable} provide a quantitative description of the QPI by
magnetic impurities in general, which include the interference between
two orthogonal helical waves discussed in Sect. \ref{standingwave} as a particular case.
The interference of helical waves corresponds to the
scattering between two TRS stationary points, like the ${\bf q}_1$'s in
Figs. \ref{fig3}(a), (b) and (c). The interesting thing is that
the LDOS in charge and spin channels from the very same pair of
TRS stationary points have quite distinct behavior. With magnetic impurities,
the power laws of charge LDOS are $R^{-2}$ and $R^{-3/2}$
for point and edge impurities respectively. As a result of TRS, the charge LDOS has higher power
indices than the $R^{-1}$ and $R^{-1/2}$ modulations of the corresponding spin-polarized LDOS,
which manifests the TRS breaking.
To distinguish the response of topological surface states to
magnetic impurities from that of the nonmagnetic impurities, \cite{Xue2009}
spin-resolved STM experiments are essential.

\subsection{Friedel oscillations for CEC with generic shape}
\label{arbitraryCEC}
In this section, we generalize the results obtained above and obtain the most general formulation of the QPI
on the surface of a 3DTI.

In the discussion of point impurity in Sect.\ref{pointimpurity}, we have focused on the case of
extremal points, around which the expansion of the CEC has nonvanishing
second derivatives. However it is in general also possible that
the principle radii of the curvature of the CEC at
the stationary points, $\rho_{xi}$,
diverges so that the third or even higher order
expansions of the CEC at the stationary points should be
employed. For example, when the stationary points are also
turning points on the CEC, see ${\bf {q}}_2^{\prime}$ in Fig.\ref{fig3}(b),
the expansion of the CEC should be kept to the third order.
In the case of edge impurity presented in Sect.\ref{edgeimpurity},
it is possible that $\rho_{xi},\rho_{xj}^{\prime}\neq 0$, but
$\rho_{xi}=\rho_{xj}^{\prime}$ so that
$P_{ij}$ diverges. This happens when the CEC near the stationary points
is highly nested, and we need to go beyond the quadratic
expansion of the CEC till the first power at which the two segments of the CEC are not nested.

To understand the LDOS behavior in ordinary and spin-resolved STM experiments
in these most general situations, we assume in general
that the first nonvanishing coefficients in the expansion of the
CEC around the stationary points have the order $l$ and
$h$ respectively, where $l,h\in {\rm N}$ are generically different.
Then $k_y(\varepsilon,k_x)$ and $k_y^{\prime}(\varepsilon^{\prime},k_x^{\prime})$
on the CEC are expanded
around the stationary points separately as $k_y=k_{yi}+\beta^{(l)}_{i}(k_x-k_{xi})^l$
and $k_y^{\prime}=k_{yj}^{\prime}+\beta_{j}^{\prime(h)}(k_x^{\prime}-k_{xj}^{\prime})^h$,
where the $\beta$'s are the first nonzero expansion
coefficients with $\beta^{(l)}_{i}=(\partial^l k_y/\partial k_{xi}^l)/l!$
and similarly for $\beta_{j}^{\prime(h)}$.
Notice that in the case of edge impurity,
if $l=h$, one more constraint $\beta^{(l)}_{i}\neq
\beta_{j}^{\prime(h)}$ should be further imposed on
the expansion to obtain a meaningful LDOS.
Having analyzed the properties of the stationary points
on the CEC, the same calculation procedures as performed in
Sects.\ref{pointimpurity} and \ref{edgeimpurity} for point
and edge impurities can be carried out in a straightforward way,
which leads to the following most general results
for point impurity
\begin{eqnarray}
\rho^{(1)}_{\mu\nu}(\omega,{\bf R})&\propto&
\frac{V}{R^{\frac{1}{l}+\frac{1}{h}}}\text{Im}\sum_{mn}\sum_{ij}\left\{\frac{e^{i(k_{yi}-k_{yj}^{\prime})R}}
{|v_{yi}v_{yj}^{\prime}||\beta^{(l)}_{xi}|^{\frac{1}{l}}|\beta^{\prime(h)}_{xj}|^{\frac{1}{h}}}\right.\nonumber\\
&\times&\left.\left[C^{\mu}_{ni,mj}C^{\nu^{\ast}}_{ni,mj}
+\frac{a^2}{|\beta^{(l)}_{xi}|^{\frac{2}{l}}R^{\frac{2}{l}}}
+\frac{a'^2}{|\beta_{xj}^{\prime(h)}|^{\frac{2}{h}}R^{\frac{2}{h}}}\right]\right\}_{\varepsilon_F},
\label{eq13}
\end{eqnarray}
and for edge impurity
\begin{eqnarray}
\rho^{(1)}_{\mu\nu}(\omega,{\bf R})
&\propto&\frac{V}{R^{\frac{1}{\text{max}(l,h)}}}\text{Im}\sum_{mn}\sum_{ij}
\left\{\frac{e^{i(k_{yi}-k_{yj}^{\prime})R}}
{|v_{yi}v_{yj}^{\prime}|}\right.\nonumber\\
&\times&|\beta^{(l)}_{xi}-\beta_{xj}^{\prime(h)}|^{\frac{-1}{\text{max}(l,h)}}
\left[C^{\mu}_{ni,mj}C^{\nu^{\ast}}_{ni,mj}\right.\nonumber\\
&+&\left.\left.(a+a')^2\left(R|\beta^{(l)}_{xi}-\beta_{xj}^{\prime(h)}|\right)^
{\frac{-2}{\text{max}(l,h)}}\right]\right\}
_{\varepsilon_F}.\label{eq14}
\end{eqnarray}
These two equations complete the key results in this work.
In the above, we have used the notation $\text{min}(l,h)$ and $\text{max}(l,h)$
to represent taking the minimum or the maximum one between $l$ and $h$.
The corresponding power laws of the Friedel oscillations
in these most general cases are summarized
in Tables. \ref{pointgeneral} and \ref{linegenral}.
We see that by taking $l=h=2$, these results recover
those exhibited in Tables. \ref{pointable} and \ref{linetable} obtained
in the last two sections.
\begin{table}[!h]
\tabcolsep 0pt \caption{General power laws of Friedel oscillations for point impurity}
\vspace*{-6pt}
\begin{center}
\def\temptablewidth{0.45\textwidth}
{\rule{\temptablewidth}{1pt}}
\begin{tabular*}{\temptablewidth}{@{\extracolsep{\fill}}ccccccc}
 &  & ordinary & spin-polarized \\   \hline
nonmagnetic&  TRP  &
$R^{-(\frac{1}{l}+\frac{1}{h})-\frac{2}{\text{min}(l,h)}}$
& -  \\
 & non-TRP & $R^{-(\frac{1}{l}+\frac{1}{h})}$ &  - \\
magnetic & TRP &
$R^{-(\frac{1}{l}+\frac{1}{h})-\frac{2}{\text{min}(l,h)}}$
  & $R^{-(\frac{1}{l}+\frac{1}{h})}$  \\
&non-TRP & $R^{-(\frac{1}{l}+\frac{1}{h})}$
&$R^{-(\frac{1}{l}+\frac{1}{h})}$
       \end{tabular*}
       {\rule{\temptablewidth}{1pt}}
       \end{center}
       \label{pointgeneral}
       \end{table}
\begin{table}[!h]
\tabcolsep 0pt \caption{General power laws of Friedel oscillations for edge impurity}
\vspace*{-6pt}
\begin{center}
\def\temptablewidth{0.42\textwidth}
{\rule{\temptablewidth}{1pt}}
\begin{tabular*}{\temptablewidth}{@{\extracolsep{\fill}}ccccccc}
 &  & ordinary & spin-polarized \\   \hline
nonmagnetic   & TRP  & $R^{-\frac{3}{\text{max}(l,h)}}$ & -  \\
 & non-TRP & $R^{-\frac{1}{\text{max}(l,h)}}$ &  - \\
magnetic& TRP &  $R^{-\frac{3}{\text{max}(l,h)}}$  & $R^{-\frac{1}{\text{max}(l,h)}}$  \\
&non-TRP & $R^{-\frac{1}{\text{max}(l,h)}}$
&$R^{-\frac{1}{\text{max}(l,h)}}$
       \end{tabular*}
       {\rule{\temptablewidth}{1pt}}
       \end{center}
       \label{linegenral}
       \end{table}

\section{Conclusions}
\label{conclusion}
In conclusion, long-distance asymptotic behavior of the LDOS for
nonmagnetic and magnetic, point and edge impurities on a generic
shape CEC are derived
in Eqs.(\ref{pointfinal}), (\ref{edgefinal}), (\ref{eq13}), and (\ref{eq14})
using the stationary phase approximation approach.
The corresponding power laws of Friedel oscillations are summarized in
Tables. \ref{pointable} to \ref{linegenral}.
The QPI induced by surface magnetic impurities is studied in particular, to illustrate the fact that
the interference patterns
of charge intensities are indistinguishable from those of nonmagnetic impurities, while the spin LDOS
show distinct behavior from those of nonmagnetic impurities.
We propose a closed ``magnetic wall'' geometry which manifests such a unique interference property of helical liquids.
These results depend only on the TRS as well as the local geometry around the
stationary points on the CEC, which provide a systematic tool for the analysis of STM experiments for generic surface states.

\acknowledgements
Q. Liu is supported by the NSFC (Grant Nos. 11004212, 11174309, and 60938004),
the STCSM (Grant Nos. 11ZR1443800 and 11JC1414500), and the
youth innovation promotion program of CAS. X. L. Qi and S. C. Zhang are supported by the Department of Energy, Office of Basic Energy
Sciences, Division of Materials Sciences and Engineering, under contract DE-AC02-76SF00515.

\end{document}